\documentclass[a4paper,12pt]{article}

\usepackage{epsfig}

\def\bc{\begin{center}}
\def\ec{\end{center}}

\def\be{\begin{equation}}
\def\ee{\end{equation}}
\def\beq{\begin{eqnarray}}
\def\eeq{\end{eqnarray}}
\def\bfig{\begin{figure}}
\def\efig{\end{figure}}
\def\bnum{\begin{enumerate}}
\def\enum{\end{enumerate}}

\begin{document}

\begin{flushright}
Journal-Ref: Astronomy Letters, 2008, v. 34, No 10, pp. 664-667
\end{flushright}

\vspace{1cm}

\bc
{\LARGE\bf Anisotropic Models of Dark Halos}\\
\vspace{0.7cm}
{\bf 
N. Ya. Sotnikova\footnotetext[1]{e-mail: nsot@astro.spbu.ru}
and 
S. A. Rodionov\footnotetext[2]{e-mail: seger@astro.spbu.ru}}\\
\vspace{0.7cm}
{\it Sobolev Astronomical Institute, \\
St. Petersburg State University,\\
Universitetskii pr. 28, 198904 Russia}
\ec

\abstract{
An iterative approach is used to construct spherically symmetric equilibrium
models with an anisotropic velocity distribution. The potentialities of the
method have been tested on models with known distribution functions, the
Osipkov--Merritt models. It is shown that models that differ significantly
from the Osipkov--Merritt models can be constructed. An $N$-body model of a
dark halo with a density distribution that approximates the results of
cosmological simulations (the Navarro--Frenk--White model) has been
constructed. The anisotropy profile has been taken to be similar to that
yielded by cosmological simulations. The constructed models can serve as
direct input data for investigating the dynamics and stability of such
systems in $N$-body simulations.}

\newpage

\section{INTRODUCTION}
\noindent
As a rule, several gravitationally significant subsystems that differ in 
structure and dynamics are identified in galaxies. Spiral galaxies are the 
most complex in this respect. They consist of a disk, a bulge, a compact 
nucleus, and an extended dark halo. Multicomponent models are required to
properly describe their dynamics. The calculations of the dynamical
evolution of composite self-consistent galactic models, including a massive
dark halo, are known to be very time-consuming. In the 1980s and 1990s,
numerical simulations with interacting galaxies (beginning with the paper by
Barnes~(1988)) were almost the only area of research in which the
computational costs were justified. The introduction of a live halo instead
of a rigid one specified analytically into the models changed radically the
outcome of the interaction. The galaxies rapidly merged together during
close collisions. This was the result of the action of dynamical friction
due to the presence of massive dark halos. 

As regards the models of isolated
galaxies (both disk and elliptical ones), a ``cheaper'' approach had long been
applied here. Usually, the dark halo potential was assumed to be rigidly
fixed. Moreover, using a live halo at a moderately large number of particles
$N$ led to inadequate dynamical evolution of the model galaxies. The stellar
disk was rapidly heated in the vertical direction (Walker~et~al.~1996). All
of the collective processes that gave rise, for instance, to transient
spirals and bars were also accelerated in the disks. Thus, for example,
Walker~et~al.~(1996) provided the growth rate of the bar-mode amplitude as a
function of the number of halo-modeling particles. In models with a rigid
halo, the bar mode did not manifest itself on time scales of 3 Gyr. A
similar behavior of the model was observed only when a very large number of
particles in the halo ($N = 450\,000$) was used. 

The potentialities of the
modern computer technology have gradually led to mass simulations with a
large number of particles and, hence, with a live halo. However, the
realization of the importance of taking into account the change in the
angular momentum of disk stars during the interaction with resonant
particles not only in the disk, but also in the live halo was more
significant. Thus, for example, the proper description of the mentioned
exchange processes prompted a revision of the views about disk stabilization
with respect to the bar mode and about the direction of evolution of the
bars in disk galaxies with a dark halo (Athanassoula~2003) that had
prevailed for thirty years (since the paper by Ostriker and Peebles~(1973)).

Cosmological simulations in which the formation of galaxies is modeled are
another area of application of self-consistent models with a dark halo. The
formation of disk galaxies is currently believed to be related to gas
accretion onto dark halos. The dark halos themselves result from
hierarchical clustering of dark matter density fluctuations (for the first 
$N$-body models of this process, see Efstathiou and Jones~1979). 

For many
dynamical problems, it is important to have simple algorithms for
constructing equilibrium models of a live halo. By this we mean the
distribution of N particles representing the system in phase space. A live
equilibrium dynamical model is fairly easy to specify if spherical symmetry
and an isotropic particle velocity distribution are assumed. The most
fruitful approach is to use a distribution function (DF) in the form $f(E)$,
where $E$ is the specific energy. According to the Jeans theorem, this ensures
that an equilibrium gravitating model will be obtained. If the DF is
unknown, but there are considerations regarding the density profile $\rho(r)$,
then $f(E)$ can be determined (analytically or numerically) via Abel's
integral relation, which is often called Eddington's formula (see, e.g.,
Binney and Tremaine 1987). Thus, for example, for the dark halo models of
great interest, the Navarro--Frenk--White model (hereafter the NFW model;
Navarro~et~al.~1996,~1997) and the Moore~et~al.~(1999) model (hereafter the
M99 model), the corresponding DFs for the isotropic case can be easily
obtained numerically (Widrow~2000; Lokas and Mammon~2001). There exist good
analytic fits to $f(E)$ (Widrow~2000). Their existence, in fact, completely
solves the problem of equilibrium initial conditions for numerical $N$-body
simulations when the isotropic NFW and M99 models are used. Another approach
is to use the Jeans equations and the assumption that the velocity
distribution is Gaussian (Hernquist~1993). However, this approach does not
guarantee an equilibrium. 

Passing to anisotropic models, even if spherically
symmetric ones, slightly complicates the preparation of numerical
simulations. In this case, a DF in the form $f(E,\,L)$, where $L$ is the
magnitude of the particle angular momentum vector, should be used to
describe an equilibrium (nonrotating) stellar system. The degree of
anisotropy of the velocity distribution described by the parameter (Binney
and Tremaine~1987) 
$\displaystyle \beta = 1 - \frac{\sigma_{\theta}^2}{\sigma_r^2}$, 
where $\sigma_{\theta}^2$ and $\sigma_r^2$ are the velocity
dispersions, respectively, in the  
$\theta$ and $r$ directions in spherical coordinates
($\beta=0$ is isotropic case and 
$\beta=1$ is the case of purely radial orbits).
In such models, $\sigma_{\theta}^2 = \sigma_{\varphi}^2$, 
where $\sigma_{\varphi}^2$ is the stellar velocity dispersion in the
$\varphi$ direction. 

The methods of finding the DF (analytically or numerically) for
anisotropic models of a special type, the so-called Osipkov­Merritt models
(hereafter the OM models; Osipkov~1979; Merritt~1985a,~1985b), are well
known. The formalism of Abel's integral transform with a slight modification
is also applicable here. Widrow~(2000) provided DF fits for the OM models
with the NFW and M99 density profiles (and for some other models).

Anisotropic spherically symmetric models can be used to investigate the
dynamical evolution of systems with a live halo. The most popular halo
models are NFW and M99. These were introduced to fit the universal density
profiles of dark halos obtained in numerical calculations within the
framework of the modern paradigm of galaxy formation through successive
mergers of cold dark matter (CDM) fluctuations. The mergers result in an
hierarchy of dark halos that differ in mass by several orders of magnitude.
The halos themselves serve as an active background against which all of the
scenarios for structure formation and galaxy formation from baryonic matter,
gas, unfold (semi-analytical models; 
White and Rees~1978; 
Fall and Efstathiou~1980; 
Blumenthal~et~al.~1986; 
Mo~et~al.~1998). 

Cosmological $N$-body simulations show that the forming halos far from the
center are not 
described by isotropic models. However, the anisotropy profile $\beta(r)$ 
is also
far from that yielded by the OM models. The generalized models by 
Cuddeford~(1991) are also unsuitable. For them, just as for the OM models, the
velocity anisotropy profile on the periphery is produced by purely radial
orbits ($\beta$ tends to 1). At the same time, for the dark halos obtained in
cosmological simulations, the degree of anisotropy is small far from the
center 
(Cole and Lacey~1996; 
Thomas~et~al.~1998; 
Colin~et~al.~2000;
Fukushige and Makino~2001; 
Diemand~et~al.~2004). 

Specifying models of
anisotropic dark halos compatible with the results of numerical simulations,
we can proceed in two ways. For example, we can take models with a small $N$
directly from cosmological simulations and use the technique of increasing
the phase density resolution (Klypin~et~al.~2001) to that required ($N$ is of
the order of several hundred thousand) in simulations with isolated galaxies
(e.g., Curir~et~al.~2007). Alternatively, we can construct $N$-body models
with a fixed anisotropy profile for a given density distribution from the
outset. 
In the latter case, one of the two recently suggested methods is
suitable. The first method is based on a generalization of the OM and
Cuddeford~(1991) method. In this method (Baes and Van~Hese~2007), a phase
density distribution function can be derived for models in which  
$\beta$ does not
tend to unity far from the center using a complex analytical technique. The
second method is based on an iterative approach to constructing equilibrium
dynamical models of galaxies and their subsystems 
(Rodionov and Sotnikova~2006; 
Rodionov and Orlov~2008). In this paper, we extend the range of
applicability of the suggested iterative method by using it in the context
of constructing models of anisotropic spherically symmetric dark halos with
a {\it fixed} velocity anisotropy profile, thereby expanding the class of
equilibrium models with prefixed properties. 

\section{THE ITERATIVE METHOD OF CONSTRUCTING EQUILIBRIUM MODELS}

\subsection{The General Idea of the Iterative Method}
\noindent
The iterative method is designed to construct nearly equilibrium $N$-body
models with a fixed density distribution (Rodionov and Sotnikova 2006;
Rodionov and Orlov 2008). It does not require any knowledge of an explicit
expression for the distribution function (DF) or the profiles of the moments
of the velocity distribution. It is only necessary to determine the type of
the system and to specify constraints on its kinematics. 

The main idea of
the iterative method is that we let the system to come to equilibrium itself
through its dynamical evolution by holding the needed parameters. The method
is implemented using the following algorithm: 

\begin{enumerate}
\item
An $N$-body model with a
fixed density distribution but with an arbitrary velocity distribution is
constructed. For example, the particle velocities may be taken to be zero.
\item
The model is allowed to evolve on a short time scale that is generally
shorter than the system crossing time. 
\item
A model with the same velocity
distribution as that in the evolved model but with the initial density
profile is constructed, i.e., the density profile is restored at the end of
each iteration. This procedure can also be considered as the transfer of the
evolving velocity distribution to the specified density model. 

If there are
initial constraints on the particle velocity distribution, then this
distribution is also "corrected" by taking into account the available
constraints\footnote{Isotropic models for which all velocity vectors should be mixed in
directions when the velocity field is transferred (Rodionov and Sotnikova
2006) are the simplest example.} 
(for a discussion, see below). 
\item
Return to item (2). 
\end{enumerate}
The
iterations are terminated when the velocity distribution ceases to change.

As a result, we obtain an $N$-body model with a fixed density profile and
fixed global kinematics that is very close to equilibrium (Rodionov and
Sotnikova 2006; Rodionov and Orlov 2008). 

In the iterative method, the
system reaches equilibrium itself in a fixed potential. Our approach differs
from previously suggested approaches, in which the system's self-adjustment
to equilibrium is also used (see, e.g., Barnes 1988), in that we forcibly
restore the required parameters and profiles to the system at the end of
each iteration. 

The idea of our iterative method is very simple. However,
when it is implemented in practice, difficulties arise at the third stage,
the stage of transferring the velocity distribution from the evolved model
to the model with the required density profile. Here, we use the velocity
distribution transfer algorithm suggested by Rodionov and Orlov (2008).

\subsection{The Velocity Distribution Transfer Algorithm}
\label{s_vtrans}

The objective of the transfer is as follows. We have the ``old'' model, a
slightly evolved model obtained at the end of the preceding iteration. We
wish to copy the velocity distribution from this model. We also have the
``new'' model constructed with a fixed density distribution. We should assign
velocities to the particles in the new model according to the velocity
distribution in the old model. 

The method for transferring the velocity
distribution from the old, evolved model to the new one was described in
detail by Rodionov and Orlov (2008). Here, we only briefly remind the
``transfer'' algorithm that we use. 

The main idea of the algorithm is to
assign exactly the same velocities to the particles in the new model as the
velocities of the particles closest to them in the old model. 

The number of
neighbors, $n_{nb}$, is an input parameter of the algorithm. For each 
particle from the old model, we introduce a variable $n_{\rm use}^j$, 
the number of times this particle is used to copy its velocity. At the 
beginning of the algorithm, we set $n_{\rm use}^j = 0$. 

First, we examine all particles in the new
model, fix the position of each such particle ${\bf r}_i$, and find the 
number of particles $n_{nb}$ from the old model that are closest to the 
position ${\bf r}_i$. Next,
we separate out a subgroup of particles with the minimum $n_{\rm use}^j$ 
from this
group and determine the particle closest to the position ${\bf r}_i$ in this
subgroup. The velocity of the particle found from the old model is assigned
to the particle under consideration from the new model. One is added to the
variable $n_{\rm use}^j$ for the particle found from the old model. 

For spherically
symmetric systems, it will suffice to seek for the nearest neighbors
according to the criterion for closeness of the radial coordinate. Thus, all
of the particles located in a thin spherical layer into which the particle
under consideration falls are considered neighbors here. This technique
improves
significantly the statistics and ensures that the constructed model will
have a spherically symmetric velocity field, i.e., the velocity distribution
will depend only on the radial coordinate. In this case, the velocities
should be ``transferred'' in a spherical coordinate system.

\subsection{Constraints on the Velocity Distribution}

\noindent
In the iterative method, the system makes a transition to equilibrium itself
in the course of its independent dynamical evolution. The direction of the
transition is not arbitrary. It is specified by forcibly fixing the density
distribution and certain kinematic parameters. If only one equilibrium model
can exist for these parameters, then one might expect the system to come
precisely to this equilibrium state in the process of iterations. Thus, for
example, when constructing isotropic spherically symmetric models, we can
maintain only the condition for equivalence of the three velocity components
without imposing any constraints on the velocity dispersion profiles. The
system itself comes to the only possible state that corresponds to the
function $f(E)$ for a given density profile (Rodionov and Sotnikova~2006).

If more than one equilibrium model can exist at fixed parameters, then an
ambiguity emerges. The question arises as to where the iterative process
will lead to in this case. One might expect that the result of the iterative
method in this case will depend on the initial conditions, while the
iterations will converge, in a sense, to the ``closest'' or distinguished
(according to particular criteria) equilibrium state. 

The problem of
nonuniqueness of the models obtained also exists for the well-known method
of Schwarzschild~(1979,~1993), which uses a library of orbits in a fixed
potential. This method was suggested to construct generally triaxial models
described via three integrals of motion. No analytical representation is
required for these integrals. It will suffice to know only the density
distribution and, accordingly, the potential of the system. However, without
a predetermined criterion for the selection of orbits in a fixed potential,
their superposition gives a whole set of models that correctly represent the
require density distribution. 

In our method, in constructing spherically
symmetric models, an ambiguity can arise when minimal constraints are
imposed on the kinematics of the system. Thus, for example, if the mean
degree of anisotropy is fixed for the entire model, then a whole set of
equilibrium anisotropic models with the same density profile can be
obtained, depending on the initial conditions on the velocity distribution.
All of the available information about the global kinematics of the system
should be used to avoid this ambiguity.

The goal of this paper is to construct equilibrium spherically symmetric
models with a fixed (generally arbitrary) anisotropy profile. To this end,
in addition to the density profile, we also held the anisotropy profile in
the process of iterations. This was achieved in the following way. At the
end of each iteration, after a readjustment of the model (with the same
velocity distribution as that in the slightly evolved model from the
preceding iteration step), we corrected the particle velocities in such a
way that the system had the required anisotropy profile. For this purpose,
we divided the model into spherical layers. Each layer contained
approximately the same number of particles. Next, we calculated the required
degree of anisotropy $\beta_k$ for each layer based on the fixed anisotropy 
profile.
The velocities of all particles in the layer were changed in such a way that
the degree of anisotropy averaged inside the layer was exactly equal to
$\beta_k$. The velocities can be corrected, for example, as follows: 
\be
\begin{array}{rcl}
v_{\varphi,\,i} &=& v^{\prime}_{\varphi,\, i} \, \sqrt{1-\beta_k} \,\,
\displaystyle\frac{\sigma^{\prime}_{r}}{\sigma^{\prime}_{\varphi}}\\
v_{\theta,\, i} &=& v^{\prime}_{\theta,\, i} \, \sqrt{1-\beta_k} \,\,
\displaystyle\frac{\sigma^{\prime}_{r}}{\sigma^{\prime}_{\theta}} \, ,
\end{array}
\ee
where $\sigma^{\prime}_{r}$, $\sigma^{\prime}_{\varphi}$ and
$\sigma^{\prime}_{\theta}$ are the current particle
velocity dispersions in the layer in the $r$, $\varphi$ and $\theta$ 
directions, respectively; 
$v^{\prime}_{\varphi,\, i}$ and $v^{\prime}_{\theta,\, i}$ are the current 
$\varphi$ and $\theta$ velocity components of particle $i$; 
$v_{\varphi,\, i}$ and $v_{\theta,\, i}$ are the corrected velocity
components.

There is some
freedom in choosing a method for correcting the particle velocity in the
layer. In the method described above, we corrected only the $\varphi$ and 
$\theta$ velocity
components. All three velocity components can be corrected in such a way
that the total change was minimal according to some chosen criterion. Test
calculations show that the result is virtually independent of the details of
the chosen procedure. 

Note also that we were interested in models with
equivalent $\varphi$ and $\theta$ velocity components, i.e., we considered 
models without rotation. To meet this condition when transferring the velocity
distribution, we ``mixed'' these velocity components. 

\section{EQUILIBRIUM ANISOTROPIC SPHERICALLY SYMMETRIC MODELS}

\noindent
Subsequently, we used the TREE algorithm
(Barnes and Hut 1986) and some of the codes from the NEMO package
(http://astro.udm.edu/nemo; Teuben 1995) to numerically solve the $N$-body
problem. All models were constructed for $N = 100\,000$. We chose the
integration step $dt$ and the softening length of the gravitational potential
$\epsilon$ in our $N$-body simulations according to recommendations from
Rodionov and~Sotnikova (2005).

The calculations were performed in a virial system of units. In this system
of units, the gravitational constant is~$G = 1$, the total mass of the model
is~$M = 1$, and the total energy of the system is~$E = -1/4$. In the NFW 
models, for which the total mass diverges, the system of units used is
specified separately.

\subsection{Test Models}

\bigskip
\noindent
{\bf Nearly homogeneous models. A Plummer sphere.} 
In stellar dynamics, there are several analytical models that are used to
test the efficiency of numerical methods. If the suggested numerical
technique completely reproduces the dynamics of known models, then it can
also be applied in more complex problems. 

In the previous section, we
described a realization of the iterative method for constructing equilibrium
spherically symmetric models with a fixed density profile and a fixed
anisotropy profile. To demonstrate the potentialities of the described
method, it is useful to turn to well-known anisotropic models for which
there are analytical expressions for the DF. The OM models are the simplest
anisotropic models. 

The anisotropy profile for all OM models is universal
irrespective of the density profile,
\be
\label{eq_om}
\beta = \frac{r^2}{r^2 + r_{\rm a}^2} \, ,
\ee
where $r_{\rm a}$ is the model parameter. This is the only parameter that
characterizes the degree of anisotropy of the models from the family in
question. The velocity distribution is predominantly isotropic inside the
sphere of radius $r_{\rm a}$ and significantly anisotropic outside the sphere 
of this
radius. Note that the degree of anisotropy for all OM models tends to unity
on the periphery of the system, i.e., radial orbits dominate on the
periphery. 

The OM models can be constructed for a density distribution
corresponding to a Plummer sphere: 
\be
\label{eq_plum}
\rho(r) = \frac{3 M}{4\pi}\frac{a^2}
{\left(r^2 + a^2\right)^{5/2}} \, ,
\ee
where $M$ is the total mass of the model and $a$ is the scale of the density
distribution (in our virial system of units, $M = 1$ and $a = 3/16$). This
density model was chosen as an example of a model with a nearly uniform
central density distribution. 

For the OM models with the density profile of
a Plummer sphere, there exists a minimum value of $r_{\rm a}/a = 3/4$
that gives a
physical model with a nonnegative DF (Osipkov~1979; Merritt~1985a). The 
$r_{\rm a}/a$ boundary that separates the physical and nonphysical models
depends on the density distribution and, in the general case (for a
particular density distribution),
can be found numerically (Carollo~et~al.~1995). However, the DF
nonnegativity for the models does not yet ensure their stability. The exact
stability boundary for the OM models can be found from numerical simulations
and, below, we demonstrate some examples related to the instability of
radial orbits. 

We used the method described in Section 2 to construct models
with the density profile~(\ref{eq_plum}) and the anisotropy
profile~(\ref{eq_om}) for several values of the parameter $r_{\rm a}$. 

\begin{figure}
\centerline{\psfig{file=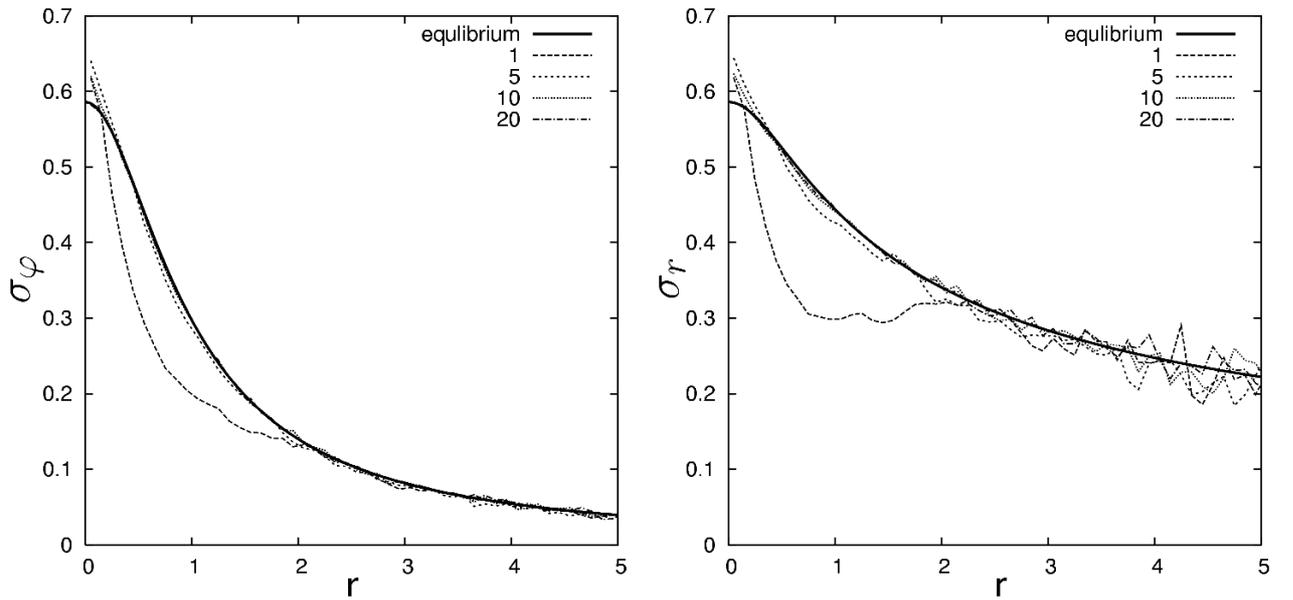,width=17cm}}
\caption{Convergence of iterations when a Plummer sphere with an anisotropy
profile corresponding to the OM model with the parameter $r_a = 0.9$ 
(the POM model) is constructed: 
(a) the $r$ dependence of the azimuthal velocity dispersion 
$\sigma_{\varphi}$; 
(b) the $r$ dependence of the radial velocity dispersion $\sigma_{r}$. 
The profiles are presented for 1, 5, 10, and 20 iterations. 
The thick solid line indicates the theoretical values of $\sigma_{\varphi}$ 
and $\sigma_{r}$ for the OM model (Merritt~1985a).
Note that the profiles for iteration 20 are virtually identical to the
theoretical ones.}
\label{fig_Pl1}
\end{figure}

The parameter of the iterative method, the time
of one iteration, was chosen to be $t_i = 10$. This choice is fairly 
arbitrary.
We can take a larger and smaller parameter $t_i$. The result of the iterative
method is virtually independent of the chosen time of one iteration
(Rodionov and Orlov~2008). The initial model for iterations was a cold model
with zero velocities. Figure~\ref{fig_Pl1} demonstrates the convergence of 
iterations
when a model with $r_{\rm a}=0.9$ is constructed (we will call it the POM 
model).
We see from Fig.~\ref{fig_Pl1} that the profiles of the stellar velocity 
dispersions in
the $r$ and  $\varphi$ directions essentially coincide with the theoretical 
profiles for
the corresponding OM model (the model with the density profile of a Plummer
sphere and with $r_{\rm a} = 0.9$) after 20 iterations. In general, all 
parameters of
the POM model, to within noise, are equal to those of the OM model. We also
experimentally checked that the POM model is an equilibrium one. In the
course of its dynamical evolution, the POM model retains all its
characteristics, to within noise, on time scales of the order of several
tens of crossing times. 

Below, we demonstrate the potentialities of the
iterative method in constructing anisotropic models differing from the OM
models. To this end, we took four anisotropy profiles that definitely do not
belong to the family~(\ref{eq_om}). 

\begin{figure}
\centerline{\psfig{file=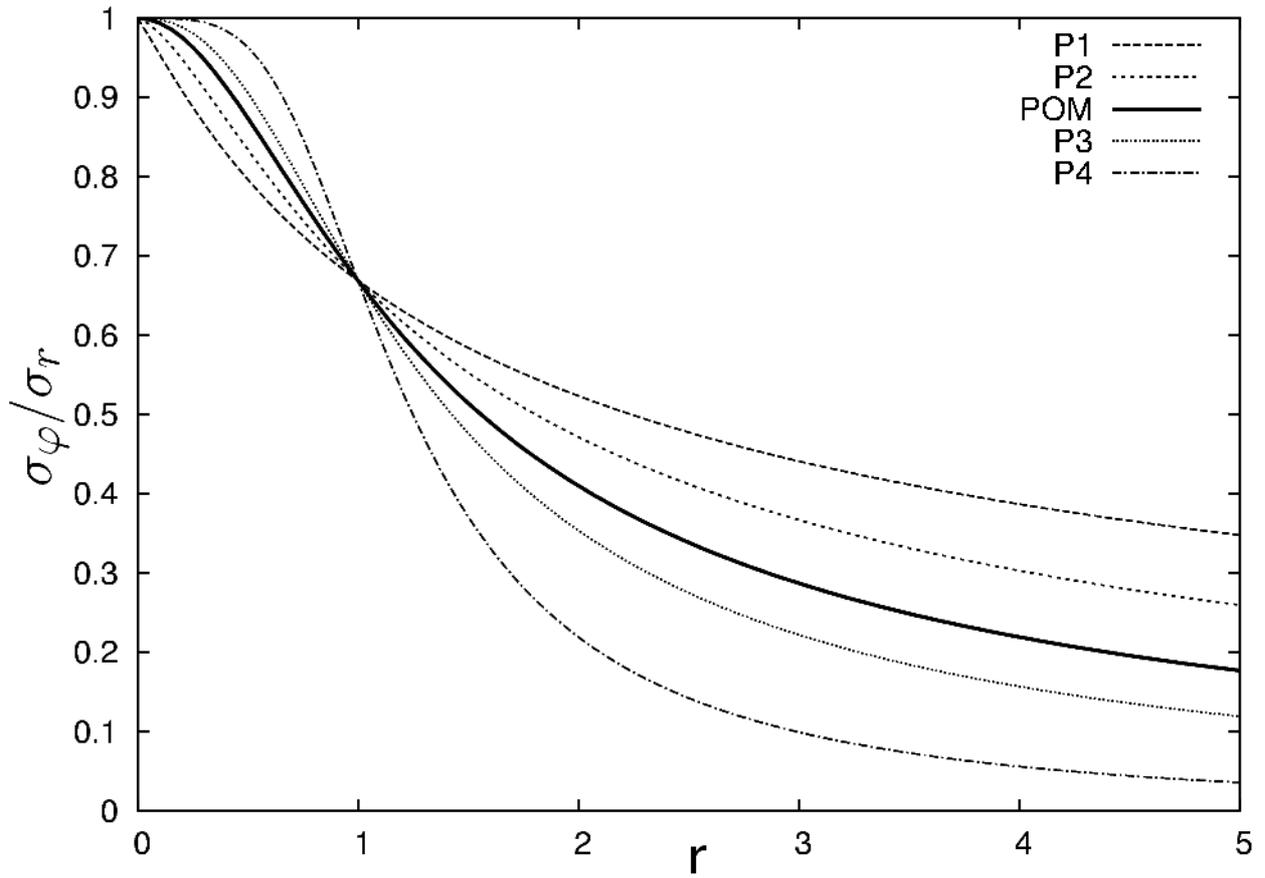,width=17cm}}
\caption{Radial profiles of the ratio $\sigma_{\varphi} / \sigma_{r}$ 
that were used to construct models P1, P2, P3, and P4 differing from the POM 
model. 
The profile of the POM model corresponds to the OM model with $r_a = 0.9$.}
\label{fig_Pl2}
\end{figure}

Figure~\ref{fig_Pl2} presents four profiles of the 
ratio\footnote{In the models under consideration, 
$\sigma_{\varphi} = \sigma_{\theta}$.}  
$\sigma_{\varphi}/\sigma_r = \sqrt{1-\beta}$. We used these profiles to 
construct models P1, P2, P3, P4, and
POM. In models P1 and P2, the $\sigma_{\varphi}/\sigma_r$ profile lies below 
and above that yielded
by the OM model in the central regions and on the periphery, respectively.
Thus, the fraction of radial orbits on the periphery decreased. We also
considered the opposite case (models P3 and P4). For these models, the
fraction of radial orbits on the periphery was higher. 

Just as in the case
of constructing the POM model, we took a cold model with zero velocities as
the initial one. The time of one iteration was $t_i = 10$. The iterations
rapidly converged (approximately in 20 steps) for all models. This suggests
that the constructed models are physical, i.e., the DF that describes the
system is nonnegative. Otherwise, the iterations would not converge.

\begin{figure}
\centerline{\psfig{file=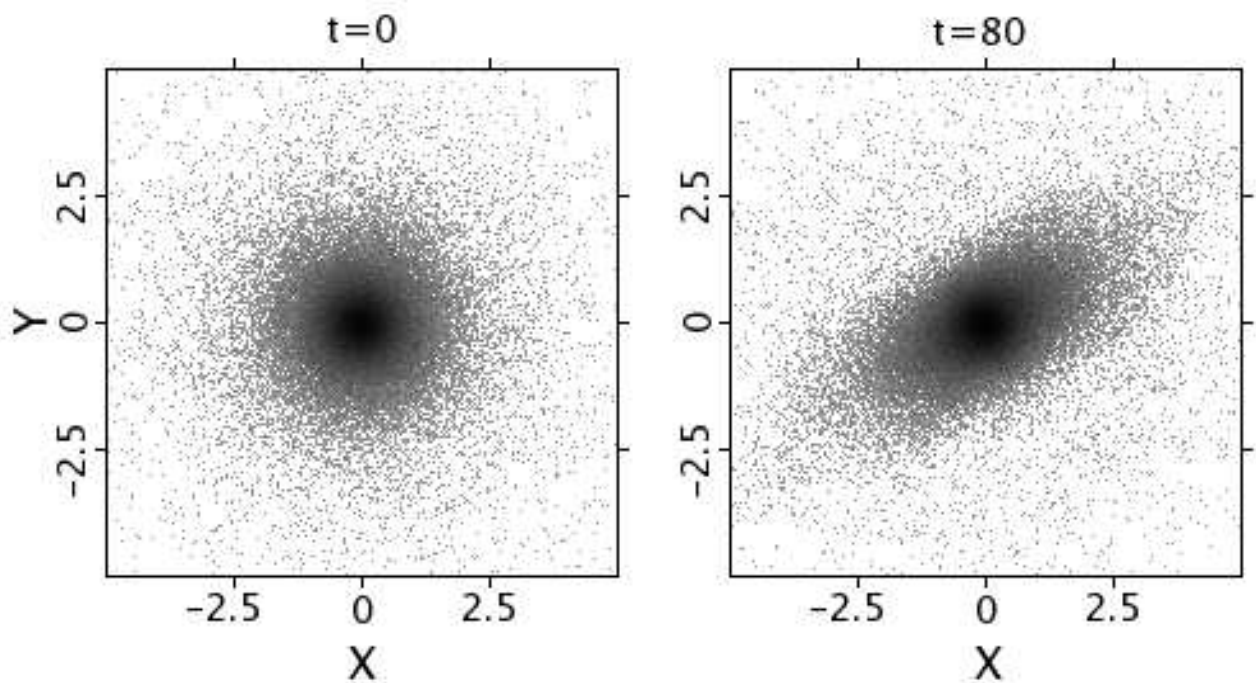,width=17cm}}
\caption{Representation of model P4 in the $x$--$y$ plane at 
(a) $t = 0$ and 
(b) $t = 80$. 
The shades of gray show the logarithm of the number of particles per
pixel. The instability of radial orbits in the outer part of the model
manifests itself as a distortion of the initial spherically symmetric model
structure.}
\label{fig_Pl3}
\end{figure}

The constructed models P1, P2, and P3 turned out to be equilibrium and
stable ones. In the course of their dynamical evolution, these models did
not change (to within noise) on a fairly long time scale, of the order of
several tens of system crossing times. They retained both the anisotropy
profiles and the remaining structural and kinematic characteristics. Model
P4 (with a wide isotropic core and a highly     
anisotropic periphery) behaved somewhat differently. This model was a
dynamically equilibrium one on a short time scale, but an instability of
radial orbits grew in the model on time scales of the order of several
crossing times ($t = 20$). This instability manifested itself as a distortion
of the initial spherically symmetric model structure on the periphery, where
nearly radial orbits dominate (Fig.~\ref{fig_Pl3}).

\bigskip
\noindent
{\bf Models with central density peaks.} 
The iterative method also easily
constructs OM-type models for other density profiles, for example, for the
Hernquist~(1990) sphere, 
\be
\label{eq_hernq}
\rho(r) = \frac{M}{2\pi}\frac{a}{r(r + a)^3}\, ,
\ee
where $M$ is the total mass of the model and $a$ is the scale of the density
profile (in our virial system of units, $M = 1$ and $a = 1/3$). The Hernquist
sphere can be used to construct models that differ significantly from the
OM-type models, for example, for models with predominantly radial orbits in
their central regions (a high degree of anisotropy). It follows from
theoretical considerations that such models are compatible only with density
profiles having central peaks (see Merritt~(1985a) for a discussion). The
density profile~(\ref{eq_hernq}) has such a
peculiarity. In contrast, the Plummer sphere considered above is an example
of a model with an almost uniform central density distribution. The central
part of this model should be isotropic, since the central radial orbits are
incompatible with the model having a homogeneous core. 

We explored the
possibilities for constructing centrally anisotropic models using the
Hernquist sphere as an example, for which the anisotropy profile was
specified in the form $\beta(r) = {\rm const}$. According to the theorem from 
An~and~Evans~(2006), for the density profile~(\ref{eq_hernq}) that 
corresponds to the Hernquist
sphere, the degree of anisotropy $\beta$in the central region cannot be 
higher
than 0.5. For a constant anisotropy profile, $\sigma_{\varphi}/\sigma_r$ 
does not depend on the
radius either. We constructed models for
$\sigma_{\varphi}/\sigma_r=0.95,\,0.9,\,0.8$, and $0.7$. We
will call these models HC0.95, HC0.9, HC0.8,
and HC0.7. These ratios $\sigma_{\varphi}/\sigma_r$ correspond to 
$\beta = 0.0975,\,0.19,\,0.36$, and $0.51$.

\begin{figure}
\centerline{\psfig{file=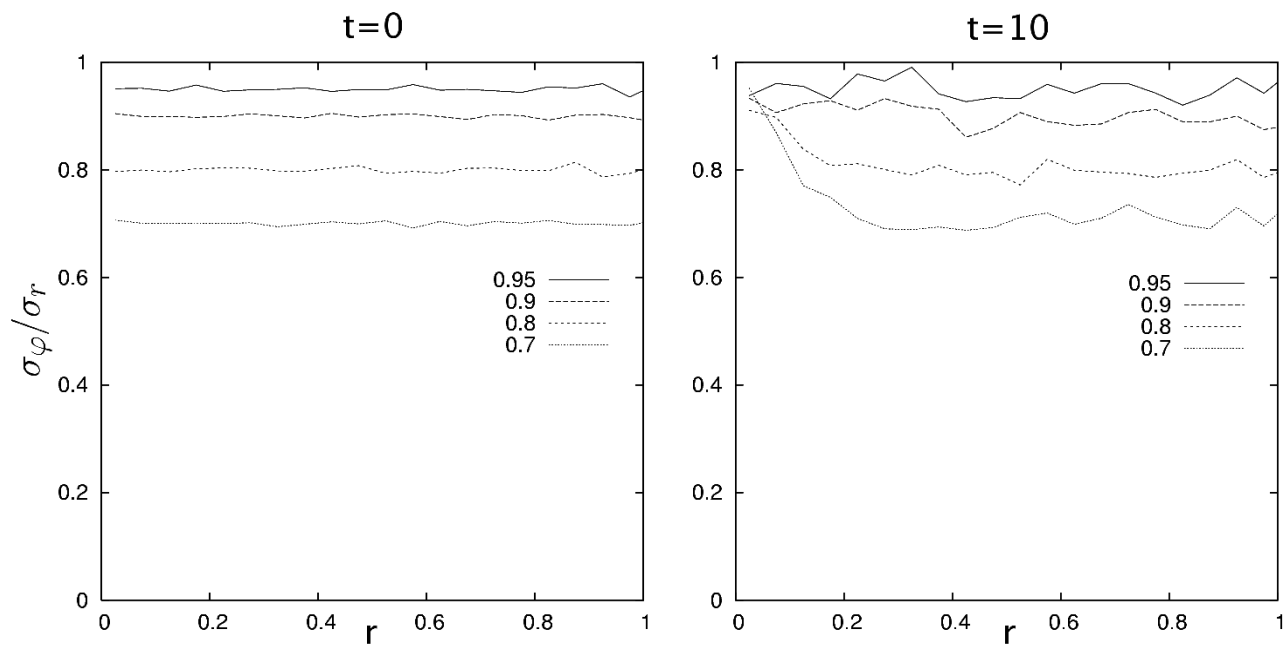,width=17cm}}
\caption{Ratio $\sigma_{\varphi} / \sigma_{r}$ versus $r$ for models 
HC0.95, HC0.9, HC0.8, and HC0.7 for two times, 
(a) $t = 0$ and 
(b) $t = 10$.}
\label{fig_Hernq}
\end{figure}

For models HC0.8 and HC0.7, the iterations converge conditionally. By the
twentieth iteration, the anisotropy profile did not change in the time of
one iteration almost over the entire length. The innermost regions
constitute an exception. The ratio $\sigma_{\varphi}/\sigma_r$ in the 
innermost regions always
reached a level approximately equal to unity. Every time we obtained an
isotropic core (Fig.~\ref{fig_Hernq}) already at the next iteration step by 
forcibly
setting a given degree of anisotropy over the entire system. Its size is
slightly smaller than the core size $a$ of the Hernquist sphere, which is 
$1/3$
in the virial system of units used. This behavior of the system could be
explained for model HC0.7c with $\sigma_{\varphi}/\sigma_r=0.7$, 
or $\beta = 0.51$. This value of $\beta$ is
higher than the above critical value of $\beta =0.5$, which is compatible 
with the
density profile of the Hernquist sphere. However, a similar behavior was
also observed for model HC0.81 with $\sigma_{\varphi}/\sigma_r=0.8$. 
The identification of a
homogeneous core probably stems from the fact that the central density peaks
compatible with radial orbits are very difficult to resolve in numerical
models. These difficulties are not related to the application of the
potential softening procedure, since the softening length for our models 
($\epsilon = 0.005$) was definitely smaller than the system core size 
$a = 1/3$. The main
reason why the central density peaks are poorly resolved is the finite
number of particles used. 

For models HC0.95 and HC0.9, the iterations
converged in the sense that the anisotropy profile ceased to change and
corresponded to the fixed profile (Fig.~\ref{fig_Hernq}). However, there was 
a slight
tendency for the core to be isotropized for these models as well. 

It should
be noted that the model with $\sigma_{\varphi}/\sigma_r=0.7$ 
(model HC0.7 in Fig.~\ref{fig_Hernq}), in addition,
turned out to be unstable and the initially spherical model turned into an
ellipsoid in the course of its dynamical evolution. 

The instability of an
anisotropic Hernquist sphere was numerically considered by Buyle et al.
(2006). In particular, they presented the results of their investigation of
models with the anisotropy profile  
$\displaystyle
\beta = \frac{\beta_0 + (r/r_{\rm a})^2}{1 + (r/r_{\rm a})^2}$
(Cuddeford~(1991) models) for various values of $\beta_0$ and $r_{\rm a}$. 
These models are
close to those we considered when $r_{\rm a} \to \infty$. As we clearly see 
from Fig.~5 in Buyle~et~al.~(2007), the parameters of our model HC0.7 are 
outside the range
of parameters for stable models. At the same time, the parameters of our
remaining models (HC0.95, HC0.9, and HC0.8) lie in the stability region,
which is also confirmed by our analysis.

\bigskip
\noindent
{\bf Anisotropic Dark Halos.}
Finally, we used our method to construct a model with a realistic anisotropy
profile, the profile obtained in cosmological simulations. 

We used the NFW
models that well describe the density profiles of cosmological dark halos.
Highresolution $N$-body simulations (Navarro et al. 1996, 1997) revealed
that the halos produced by the gravitational clustering of dark matter have
a universal structure. The properly scaled density distribution does not
depend on the applied cosmological model. It is represented by the following
simple formula: 
\be
\rho(r) = \frac{4 \rho_{\rm s}}{(r/r_{\rm s})(1+r/r_{\rm s})^2} \, ,
\label{eq_NFW}
\ee
where $r_{\rm s}$ is the
characteristic radius at which the logarithmic slope of the density profile
is $d \ln \rho / d \ln r = -2$ and $\rho(r_{\rm s}) = \rho_{\rm s}$. 
The shape of the density profile is
characterized by the concentration parameter $c=R_{\rm vir}/r_{\rm s}$. 
Here, $R_{\rm vir}$ is
the so-called virial radius. It is introduced in such a way that the mean
density inside the sphere of radius $R_{\rm vir}$ is higher than the critical 
density specified in cosmological models by a fixed factor 
($\delta_{\rm vir}$). In various
papers, this radius is assumed to be $\Delta_{\rm vir}=100,\,178,\,200$. 

We will use a
system of units with $G = 1$, $r_{\rm s}=1$, and 
$rho_{\rm s}=1/4$. Fixing $r_{\rm s}$ means that the
virial radius can be expressed in terms of the concentration parameter.
Thus, for example, the concentration parameter $c = 10$ commonly used in
galaxy formation models gives the virial radius $R_{\rm vir}=10$. 

The density profile~(\ref{eq_NFW}) leads to a diverging mass. Therefore, in
$N$-body models, it 
should be cut off at some radius $R_{\rm cut}$. To avoid the boundary 
effects, we
assumed in our calculations that $R_{\rm cut}=40\,r_{\rm s}$. Starting from 
$r_{\rm cut}=20$, we represented the density profile by a spline. 
At $c = 10$, this gives $r_{\rm cut}=2\,R_{\rm vir}$ and 
$R_{\rm cut}=4\,R_{\rm vir}$. 

In $N$-body simulations, the dark matter halos have
an anisotropic velocity distribution. The anisotropy profile $\beta(r)$ 
usually
has such a structure that $\beta(r) \approx 0$ within the sphere of a radius 
comparable to $r_{\rm s}$. Thus, the dark halo model has an isotropic core. 
Further out, $\beta(r)$ gradually increases, reaching  
$\beta(r) = 0.35-0.5$ at $r=R_{\rm vir}$. In terms of the
ratio $\sigma_{\varphi}/\sigma_r$, this gives
$\sigma_{\varphi}/\sigma_r=0.7-0.8$ 
(Cole and Lacey~1996; 
Thomas~et~al.~1998; 
Colin~et~al.~2000; 
Fukushige and Makino~2001; 
Diemand~et~al.~2004).

In our calculations, we took a $\sigma_{\varphi}/\sigma_r$ profile in the 
form  
\be
\frac{\sigma_{\varphi}}{\sigma_r} =
\frac{0.2}{\displaystyle\sqrt{\left(\frac{r}{0.9}\right)^2+1}} + 0.8 \, .
\label{eq_anfw}
\ee
In this case, the ratio $\sigma_{\varphi}/\sigma_r$ is close to unity within 
the core of radius $r_{\rm s}=1$ and decreases to 0.8 at the distance 
$r=R_{\rm vir}$. 

Using the iterative
method, we constructed a model with the density profile~(\ref{eq_NFW}) and an
anisotropy profile corresponding to~(\ref{eq_anfw}). The parameter of the 
iterative
method, the time of one iteration, was chosen to be $t_i = 10$. The initial
model for iterations was a cold model with zero velocities. As in the case
of constructing the Plummer and Hernquist models described above, the
iterations converged rapidly (approximately in 20 steps). 

\begin{figure}
\centerline{\psfig{file=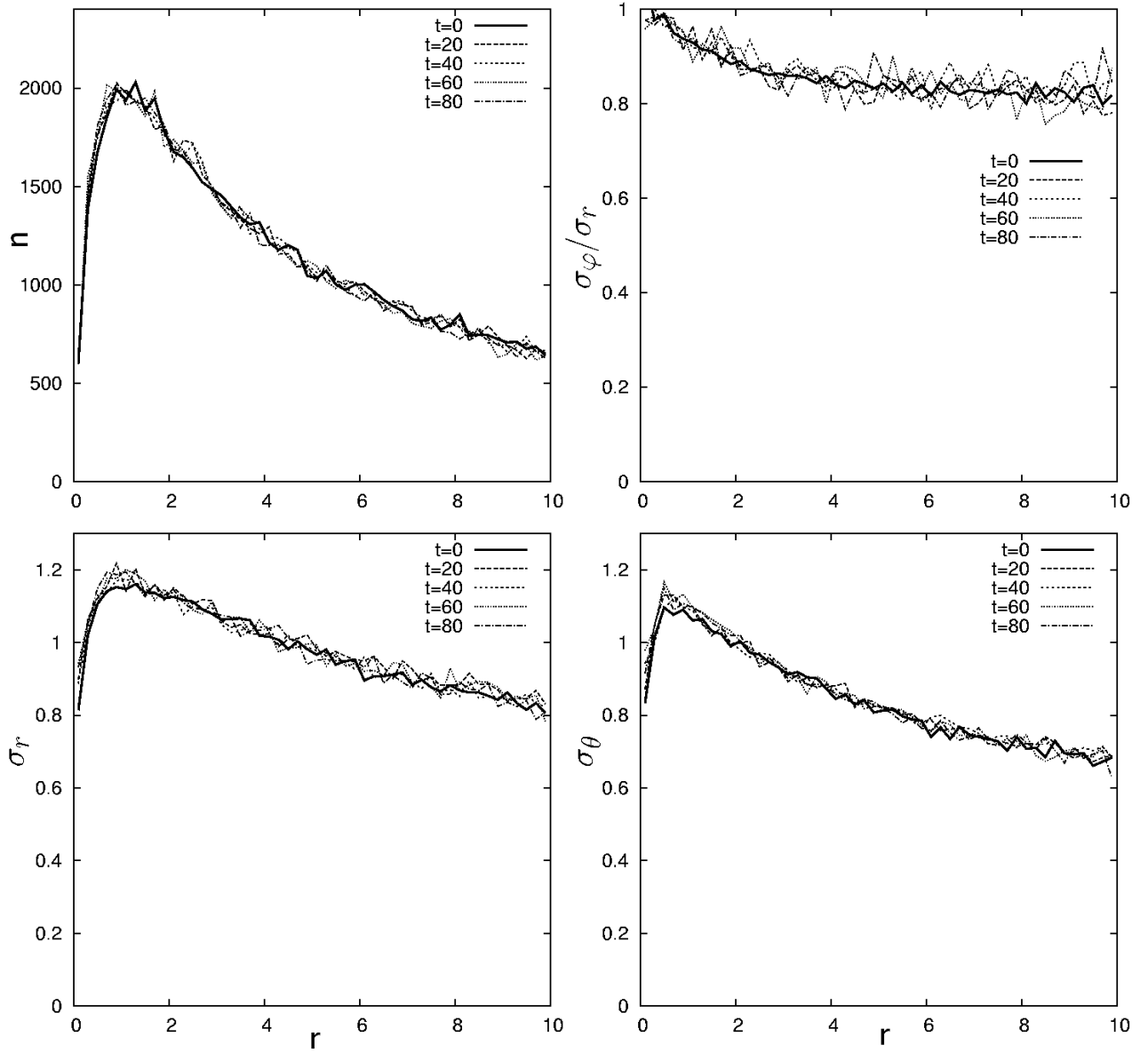,width=17cm}}
\caption{Equilibrium test for the ANFW model. 
For several times ($t$ = 0, 20, 40, 60, and 80), the radial profiles are 
presented for the following quantities: 
$n$ is the number of particles in concentric spherical layers,  
$\sigma_{\varphi} / \sigma_{r}$, $\sigma_{r}$ and $\sigma_{\theta}$.}
\label{fig_NFW}
\end{figure}

Figure~\ref{fig_NFW} shows the
results of our equilibrium test for the constructed model (ANFW). We see
that the structural and kinematic profiles of the model are well preserved
on a time scale comparable to several system crossing times. 

To demonstrate
the potentialities of the suggested iterative method, we constructed only
one anisotropic NFW model. It turned out to be stable. However, the
stability of anisotropic halo models with the density profile~(\ref{eq_NFW}) 
has not
been studied in the literature even for the simplest types of anisotropy
like the OM models. Our technique allows halo models to be constructed in a
broad class of anisotropy profiles. In the future, we are planning to study
the stability of such models in more detail.

\section{DISCUSSION AND CONCLUSIONS}

\noindent
The main advantage of the method used here is that we do not need to know an
explicit expression for the DF to construct a model of the stellar system.
It will suffice only to ascertain what number of integrals of motion it
depends on. This determines the type of the equilibrium dynamical models
being realized. Subsequently, the model dynamically adjusts itself to the
specified constraints on the kinematics of the system. 

Our method can be
compared to the well-known method of Schwarzschild~(1979) and its
modification known as M2M, made-to-measure (Syer and Tremaine~1996; 
De~Lorenzi~et~al.~2007). In Schwarzschild's and M2M methods, the system is
constructed as a superposition of particles moving, as a rule, in a fixed
potential. The weights with which the particles contribute to the density
and velocity distributions can be determined by minimizing the deviations of
various parameters of the model being constructed from those of the target
model. Our method is simpler: the system adjusts itself to equilibrium. We
only hold a number of fixed parameters. In the realization of the iterative
method described here, we adjust the velocities to fixed constraints when
transferring the velocity distribution from one iteration step to another.

The question arises as to how general these constraints should be and
whether more detailed kinematics gives rise to systems that are actually not
realizable. 

For anisotropic spherically symmetric models, we can fix the
velocity anisotropy profile $\beta(r)$ as a constrain on the kinematics of 
the system. An alternative constraint is to consider the mean degree of
anisotropy of the system or the mean value of $\sigma_{\varphi}/\sigma_r$ 
for the entire system. 

In
this paper, we described spherically symmetric models for which the
anisotropy profile was fixed from the outset. Let us make several remarks
concerning the second variant of constraints. 

It seemed to us that global
constraints on the kinematics of the system could give some of the selected
models. Thus, for example, Rodionov and Orlov~(2008) used the iterative
approach to construct models with a completely different geometry, stellar
disks. Fixing the total angular momentum of the disk as a global parameter
of the system yielded a oneparameter family of models. Their characteristic
feature was a Gaussian velocity distribution in three directions. For
spherically symmetric anisotropic models, all turned out to be more complex.
If the mean value of $\sigma_{\varphi}/\sigma_r$ is fixed for the entire 
system, but the profile $\beta(r)$ is not maintained forcibly, then the 
solution is found to be ambiguous. It
depends strongly on the iteration parameters (the initial velocities, the
time of the system's ``independent'' dynamical evolution during one iteration,
etc.). Thus, in the class of anisotropic models under consideration, the
approach with global constraints on the kinematics of the system does not
set apart any family of finite models. 

Of course, our approach does not
allow equilibrium anisotropic models with an arbitrary anisotropy profile to
be constructed. This is possible only for $\beta(r)$ that are compatible with 
a nonnegative DF. If the DF $f(E,\,L)$ corresponding to spherically symmetric
anisotropic models can take on negative values at admissible $E$ and $L$, then
the iterative method cannot be applied to the construction of such systems.
For example, no equilibrium model of a Plummer sphere with the OM anisotropy
profile~(\ref{eq_om}) and $r_a/a < 3/4$ can be constructed by the iterative
method.

The
constructed equilibrium models may turn out to be unstable (e.g., models P4
and HC0.7 described here). The stability criteria in stellar dynamics impose
the most stringent constraints on the global parameters of real systems. In
the context of the problem of constructing anisotropic spherically symmetric
models, our method gives direct material for investigating the stability of
such models and constraints on the possible shape of the anisotropy profile
$\beta(r)$. Our method expands significantly the class of stable
equilibrium models. For a given density distribution, this makes it possible
to draw a boundary over the actually realizable anisotropy profiles. 

The
question of the stability of anisotropic models is particularly topical as
applied to dark halo models with the NFW or M99 density distribution.
Whereas the family of so-called 
$\gamma$-models\footnote{The Hernquist sphere belongs to this family.} 
(see, e.g., Dehnen~1993) was
analyzed for stability in sufficient detail for various anisotropy profiles
(Carollo~et~al.~1995; Meza and Zamorano~1997; Buyle~et~al.~2006), the
anisotropic NFW and M99 models are virtually unstudied. Since we now have a
tool for constructing such models in our hands, we are planning to return to
this question. 

It should also be noted that the constructed anisotropic dark
halo models with a density distribution corresponding to cosmological dark
halos can serve as direct input data for investigating the dynamics of such
systems in $N$-body simulations. It is easy to generalize the models
described here to spherically symmetric anisotropic models with rotation.
These models can be used in the problems of gas accretion onto dark halos
and in modeling the formation of disk galaxies. Comparison of the global
parameters for stellar disks at various stages of their formation with
observational data for disk galaxies at various redshifts can give an
estimate of the relative efficiency of the external and internal processes
that affect the evolution of galaxies. Another result of solving this
problem can be the verification of several relations between structural and
kinematic parameters of the disk and the halo that follow from the
well-known semi-analytical models for the formation of disk galaxies 
(White and Rees~1978; 
Fall and Efstathiou~1980; 
Blumenthal~et~al.~1986; 
Mo~et~al.~1998).

\bc
ACKNOWLEDGMENTS 
\ec

This work was supported by the Russian Foundation for
Basic Research (project no. 06-02-016459) and a grant for support of leading
scientific schools from the President of Russia (NSh-8542.2006.2).

\bc
REFERENCES                           
\ec

\noindent
1. J. H. An and N. W. Evans, Astron. J. 131, 782 (2006).

\noindent
2. E. Athanassoula, Mon. Not. R. Astron. Soc. 341, 1179 (2003). 

\noindent
3. M. Baes and E. Van Hese, Astron. Astrophys. 471, 419 (2007). 

\noindent
4. J. Barnes, Astrophys. J. 331, 699 (1988). 

\noindent
5. J. Barnes and P. Hut, Nature 324, 446 (1986). 

\noindent
6. J. Binney and S. Tremaine, Galactic Dynamics 
(Princeton Univ. Press, Princeton, 1987).   

\noindent
7. G. R. Blumenthal, S. M. Faber, R. Flores, and J. R. Primack, 
Astrophys. J. 301, 27 (1986). 

\noindent
8. P. Buyle, E. Van Hese, S. De Rijcke, and H. Dejonghe, 
Mon. Not. R. Astron. Soc. 375, 1157 (2006). 

\noindent
9. C. M. Carollo, P. T. de Zeeuw, and R. P. van der Marel, 
Mon. Not. R. Astron. Soc. 276, 1131 (1995). 

\noindent
10. Sh. Cole and C. Lacey, 
Mon. Not. R. Astron. Soc. 281, 716 (1996). 

\noindent
11. P. Coli´ n, A. A. Klypin, and A. V. Kravtsov, 
Astrophys. J. 539, 561 (2000). 

\noindent
12. P. Cuddeford, Mon. Not. R. Astron. Soc. 253, 414 (1991). 

\noindent
13. A. Curir, P. Mazzei, and G. Murante, 
Astron. Astrophys. 467, 509 (2007). 

\noindent
14. F. De Lorenzi, V. P. Debattista, O. Gerhard, and N. Sambhus, 
Mon. Not. R. Astron. Soc. 376, 71 (2007). 

\noindent
15. W. Dehnen, Mon. Not. R. Astron. Soc. 265, 250 (1993). 

\noindent
16. J. Diemand, B. Moore, and J. Stadel, 
Mon. Not. R. Astron. Soc. 352, 535 (2004). 

\noindent
17. G. Efstathiou and B. J. T. Jones, 
Mon. Not. R. Astron. Soc. 186, 133 (1979). 

\noindent
18. S. M. Fall and G. Efstathiou, 
Mon. Not. R. Astron. Soc. 193, 189 (1980). 

\noindent
19. N. Fukushige and J. Makino, 
Astrophys. J. 557, 533 (2001). 

\noindent
20. L. Hernquist, Astrophys. J. 356, 359 (1990). 

\noindent
21. L. Hernquist, Astron. Astrophys. Suppl. Ser. 86, 389 (1993). 

\noindent
22. A. Klypin, A. Kravtsov, J. Bullock, and J. Primack, 
Astrophys. J. 554, 903 (2001). 

\noindent
23. E. L. Lokas and G. A. Mammon, 
Mon. Not. R. Astron. Soc. 321, 155 (2001). 

\noindent
24. A. Meza and N. Zamorano, Astrophys. J. 490, 136 (1997).

\noindent
25. D. Merritt, Astron. J. 90, 1027 (1985a). 

\noindent
26. D. Merritt, Mon. Not. R. Astron. Soc. 214, 25 (1985b). 

\noindent
27. H. J. Mo, S. Mao, and S. D. M. White, 
Mon. Not. R. Astron. Soc. 295, 319 (1998). 

\noindent
28. B. Moore, T. Quinn, F. Governato, J. Stadel, and G. Lake, 
Mon. Not. R. Astron. Soc. 310, 1147 (1999). 

\noindent
29. J. F. Navarro, C. S. Frenk, and S. D. M. White, 
Astrophys. J. 462, 563 (1996). 

\noindent
30. J. F. Navarro, C. S. Frenk, and S. D. M. White, 
Astrophys. J. 490, 493 (1997). 

\noindent
31. L. P. Osipkov, Pis'ma Astron. Zh. 5, 77 (1979) 
[Sov. Astron. Lett. 5, 42 (1979)]. 

\noindent
32. J. P. Ostriker and P. J. E. Peebles, 
Astrophys. J. 186, 467 (1973). 

\noindent
33. S. A. Rodionov and V. V. Orlov, 
Mon. Not. R. Astron. Soc. 385, 200 (2008). 

\noindent
34. S. A. Rodionov and N. Ya. Sotnikova, 
Astron. Zh. 82, 527 (2005) [Astron. Rep. 49, 470 (2005)]. 

\noindent
35. S. A. Rodionov and N. Ya. Sotnikova, 
Astron. Zh. 83, 1091 (2006) [Astron. Rep. 50, 983 (2006)]. 

\noindent
36. M. Schwarzschild, Astrophys. J. 232, 236 (1979). 

\noindent
37. M. Schwarzschild, Astrophys. J. 409, 563 (1993). 

\noindent
38. D. Syer and S. Tremaine, Mon. Not. R. Astron. Soc. 282, 223 (1996). 

\noindent
39. P. J. Teuben, ASP Conf. Ser. 77, 398 (1995). 

\noindent
40. P. A. Thomas, J. M. Colberg, H. M. P. Couchman, et al., 
Mon. Not. R. Astron. Soc. 296, 1061 (1998). 

\noindent
41. I. R. Walker, C. Mihos, and L. Hernquist, 
Astrophys. J. 460, 121 (1996). 

\noindent
42. S. D. M. White and M. J. Rees, 
Mon. Not. R. Astron. Soc. 183, 341 (1978). 

\noindent
43. L. M. Widrow, Astrophys. J. Suppl. Ser. 131, 39 (2000).

\flushright{\it Translated by V. Astakhov}

\end{document}